\documentclass[5p,twocolumn,times,number]{elsart5p}
\usepackage{ae}
\usepackage{amsmath}
\usepackage{amssymb}
\usepackage{graphics}
\usepackage{graphicx}
\usepackage{epsfig}
\usepackage{subfigure}
\usepackage{dcolumn}
\usepackage{bm}
\usepackage{lineno}

\linenumbers
\begin{document}
\begin{frontmatter}

\title{Performances of silicone coated high resistive bakelite RPC}
\author{S. Biswas\thanksref{label1}\corauthref{cor1}},
\ead{saikatb@veccal.ernet.in}
\thanks[label1]{}
\corauth[cor1]{}
\author[label2]{Purba Bhattacharya},
\author[label2]{S. Bhattacharya},
\author[label2]{S. Bose},
\author[label1]{S. Chattopadhyay},
\author[label2]{N. Majumdar},
\author[label2]{S. Mukhopadhyay},
\author[label2]{S. Saha},
\author[label1]{Y.P. Viyogi}

\address[label1]{Variable Energy Cyclotron Centre, 1/AF Bidhan
Nagar, Kolkata-700 064, India}
\address[label2]{Saha Institute of Nuclear Physics, 1/AF Bidhan Nagar, Kolkata-700
064, India}

\begin{abstract}
Performances of several single gap (gas gap 2~mm) prototype Resistive Plate Chambers (RPC)
made of high resistive ($\rho$ $\sim$ 10$^{10}$ - 10$^{12}$ $\Omega$ cm) bakelite,
commercially available in India have been studied in recent times. To make the inner
electrode surfaces smooth, a thin coating of silicone has been applied. An efficiency >~90\% and
time resolution $\sim$ 2 ns (FWHM) have been obtained for both the streamer and the avalanche mode. The induced charge distributions of those silicone coated RPC are studied and the results are presented. A numerical study on the effect of surface roughness of the resistive electrodes on the electric field of the device has been carried out using Garfield-neBEM code. A few results
for a simplified model representing surface roughness, measured using a surface profilometer for the bakelite surfaces, have also been presented.
\end{abstract}

\end{frontmatter}

\section{Introduction}
\label{}
High resistive ($\rho$ $\sim$ 10$^{10}$ - 10$^{12}$ $\Omega$ cm) bakelite RPCs \cite{RSRC81},
prepared with a silicone compound coating on the inner electrode surfaces,
were studied in a cosmic ray test bench to characterize its long term stability of operation \cite{SB109}. Need for the silicone compound coating had been illustrated through the measurement of variation of efficiency with applied high voltage (HV) in our earlier work \cite{SB109}. Study of variation of time resolution and cross-talk between neighbouring pick-up strips with applied HV were also reported earlier \cite{SB309,SB209}. Such high
resistive electrode-based RPCs are being explored as active elements of a large scale iron calorimeter for the proposed India-based Neutrino Observatory (INO) \cite{INO06}.

One of the limitations of the streamer mode operation of high resistive bakelite RPCs is that
the rate handling capability is relatively worse than that in the avalanche mode. For the same energetic particle falling on the RPC, the charge content contributing to the resulting pulses
in the two modes differ by orders of magnitude. This, as well as the physical extent of the streamer and the electron avalanche, may result in variation of the pulse timing properties.
These aspects need to be studied both through experiment and through simulation.

Stability of operation of the RPCs and its dependence on the detector parameters are the other important aspects which need to be studied in detail. It is established that a rough inner electrode surface of an RPC is prone to cause field emission, which is a source of high dark
current \cite{CLu09}. In this context, it is relevant to investigate the effect of surface roughness on the detector dynamics. It is conceivable that imperfections on the bakelite surface may be responsible to cause local non-uniformity of the electric field. Depending upon their shape and size, they may often lead to discharge severely degrading the performance of the device.

In this article, some experimental results on operation of our bakelite RPC in the avalanche mode, systematic studies and comparison of charge contents of the resulting pulses in the avalanche mode and the streamer mode operation of the same RPCs are reported. In our endeavor of corroboration of experimental results with simulation, preliminary results of numerical simulation of the effect of roughness of the resistive bakelites on the field configuration are also reported. Using Garfield-neBEM \cite{Garfield,neBEM} code, the field calculation has been done for several models designed on the basis of measurements of the bakelite surface profiles \cite{SB109}.

\section{Test setup}
\label{}

All the RPCs were tested in a cosmic ray test bench described in Ref.~\cite{SB109}.
The cosmic ray telescope was constructed using three scintillators, two placed above the RPC
and one below. Triple coincidence of the signals, obtained from these three scintillators was used to select a cosmic ray event(master trigger). The signals obtained from the pick-up strips were put in coincidence with the master trigger obtained as above. This is referred to as the coincidence trigger. Finally, the efficiency was calculated as the ratio of the coincidence trigger to the master trigger.

Premixed gas of argon, isobutane and tetrafluroethane(R-134a) in the volume ratio of
55/7.5/37.5 (equivalent to a mass mixing ratio of 34/7/59) was used in the streamer mode,
while in the avalanche mode, R-134a/isobutane in 95/5, R-134a/isobutane/SF$_6$ in 95/4.5/0.5 and 95/2.5/2.5 volume mixing ratio were used. A typical flow rate of 0.4 ml per minute was maintained by the gas delivery system \cite{SBose109}, resulting in $\sim$ 3 changes of gas gap volume per day.

\begin{figure}
\includegraphics[scale=0.5]{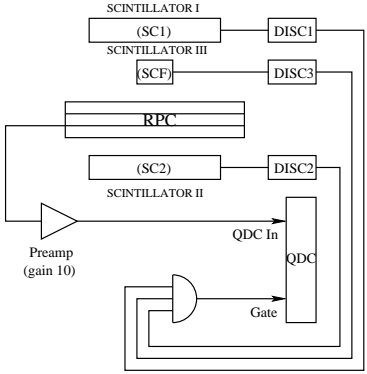}\\
\caption{\label{fig:epsart}Schematic representation of the charge spectrum
measurement setup. Phillips Scientific 7186 Charge to Digital Converter (QDC) was used.} \label{fig:1}
\end{figure}

Charge content of the pulses were measured using a charge to digital converter, referred to as QDC. The set-up is shown in Fig.~\ref{fig:1}, which is very similar to that used in Ref.~\cite{CLu06}. The master trigger gate of width 1.5 $\mu$s was also used as gate to the QDC. It was observed that the charge spectrum and the measured mean charge were independent of gate width over a range of 150 ns to 1.5 $\mu$s. The pulse height in the streamer mode was noted to be a few hundred mV, while that in the avalanche mode was found to be $<$~10 mV.  A $10\times$ voltage pre-amplifier was used for the avalanche mode operation of the RPCs.

\section{Results}
\label{}

In the avalanche mode, the RPCs were tested using different gas mixtures mentioned as above.
The efficiency and time resolution of the RPCs were studied by varying the applied HV. The efficiency plateau was obtained at >~90\% for all the gas mixtures, and it was found to be marginally higher for gas mixtures containing less amount of SF$_6$. At the plateau region, the time resolution was found to be $\sim$ 2.5 ns(FWHM) and the average signal arrival time
decreased with the increase of HV, which is a common feature of any gas filled detector.

\begin{figure}[htb]
\includegraphics[scale=0.3]{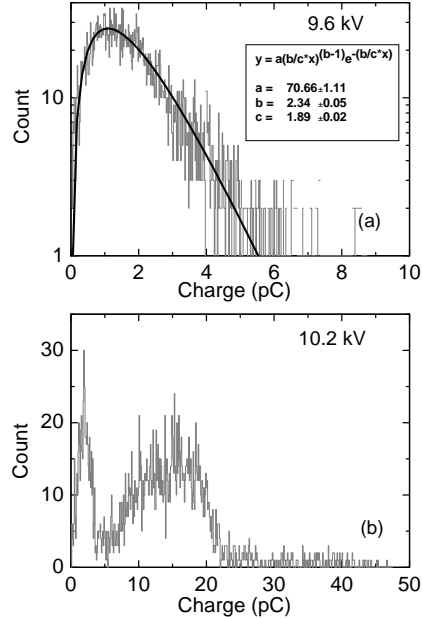}\\
\caption{\label{fig:epsart}The charge spectrum of the RPC operated in the avalanche mode using the gas mixture of R-134a/isobutane/SF$_6$ in 95/4.5/0.5 ratio, at two different high voltages.}\label{fig:2}
\end{figure}

\label{}
\begin{figure}[htb]
\includegraphics[scale=0.23]{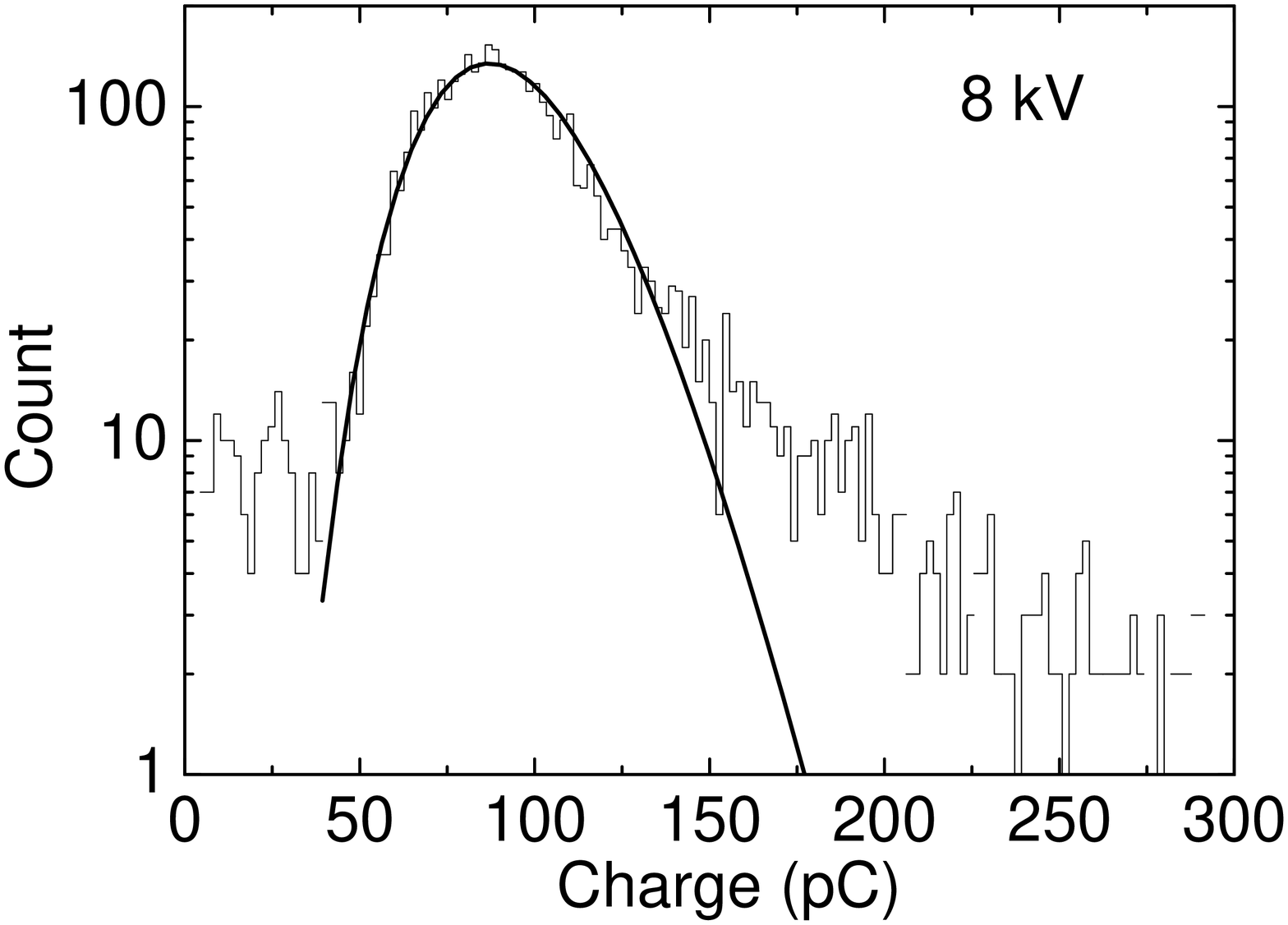}\\
\caption{\label{fig:epsart} The charge distribution spectrum in the streamer mode with a gas mixture of Argon/isobutane/R-134a in 55/7.5/37.5 ratio.}\label{fig:3}
\end{figure}

In Fig.~\ref{fig:2}~(a), a typical charge spectrum of the RPC operated in the avalanche mode
(HV~$=9.6$~kV) using the gas mixture of R-134a/isobutane/SF$_6$ in 95/4.5/0.5 ratio, together with the Polya distribution fitting curve is shown. At higher HV values, a streamer peak appears after the avalanche as shown in Fig.~\ref{fig:2}~(b). The charge distribution spectrum in the streamer mode with a gas mixture of Ar, isobutane and R-134a in 55/7.5/37.5 ratio is shown in Fig.~\ref{fig:3}. Corresponding best-fit Polya distribution curve to the streamer charge
spectrum is also shown in the same plot. It is clear that the fitted curve underpredicts the pulse charge content at the higher charge tail of the spectrum. This possibly indicates the onset of a breakdown mechanism responsible for the excess streamer charge beyond that can be predicted by the clustering model.

\begin{figure}
\includegraphics[scale=0.3]{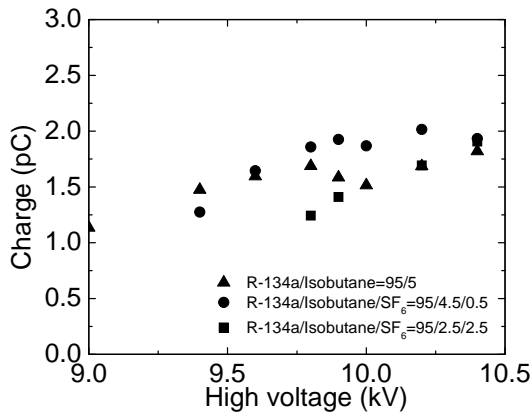}\\
\caption{\label{fig:epsart} The induced charge as a
function of HV for a silicone coated RPC in the avalanche
mode using three different gas mixtures.}\label{fig:4}
\end{figure}

\begin{figure}
\includegraphics[scale=0.3]{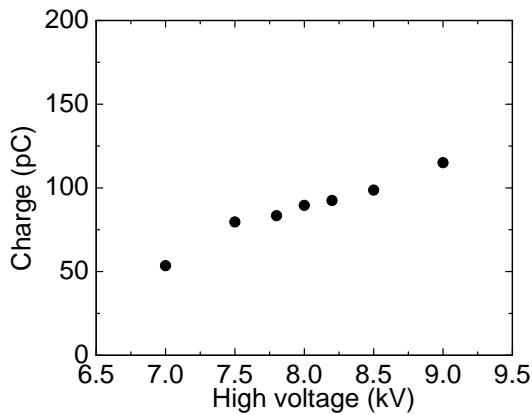}\\
\caption{\label{fig:epsart} The induced charge as a
function of HV for a silicone coated RPC in the streamer mode with a gas mixture of Argon/isobutane/R-134a in 55/7.5/37.5 ratio.}\label{fig:5}
\end{figure}

The induced charge as a function of the applied HV in the avalanche mode is shown in
Fig.~\ref{fig:4}, whereas that in the streamer mode is shown in Fig.~\ref{fig:5}. For the RPC, when operated in the avalanche mode with the gas mixtures of R-134a/isobutane in 95/5 and R-134a/isobutane/SF$_6$ in 95/4.5/0.5 ratio it was observed that at higher voltages a substantial streamer peak appears. But the charge shown in Fig.~\ref{fig:4} is the mean charge of the avalanche peak only. It is clear from the figures that the charge saturates at $\sim$ 1-2 pC in the avalanche mode for all the gas compositions, while in the streamer mode, the charge is $\sim$ 100 pC.

\section{Surface roughness and its effect on the electric field}
\label{}
\begin{figure}[htb]
\begin{center}
\includegraphics[scale=0.4]{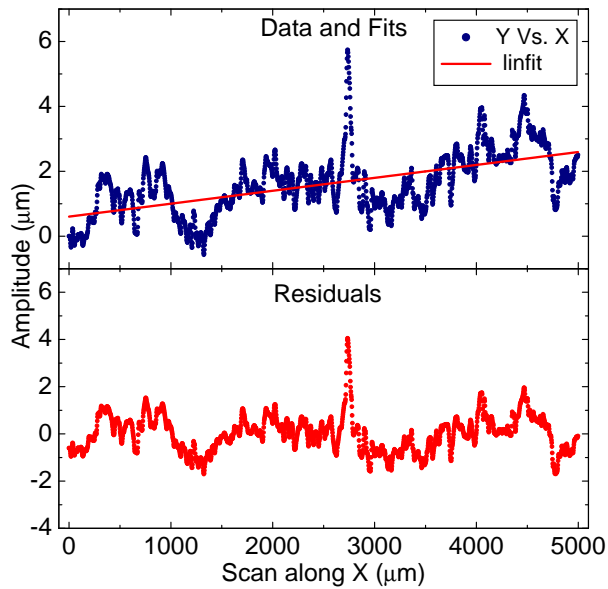}\\
\caption{\label{fig:linfit} A typical surface profile of bakelite P120 measured with DEKTAK 117
profilometer. The least square fit to the data and the residuals w.r.t the fit are shown.}
\label{fig:surfacedata}
\end{center}
\end{figure}

\begin{figure}[hb]
\begin{center}
{\label{Set3c}\includegraphics[scale=0.75]{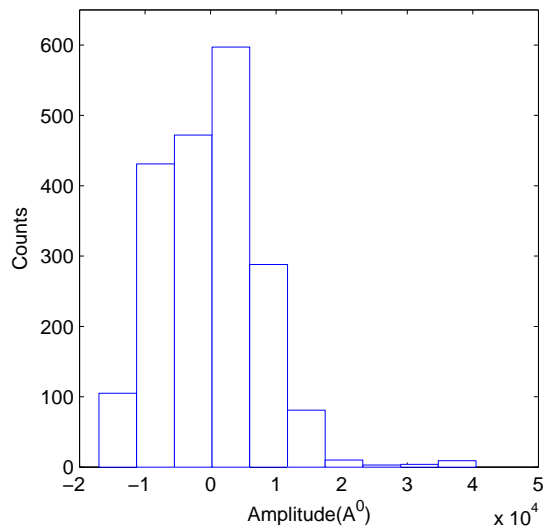}}
\caption{Histogram of amplitudes with a maximum of 4 $\mu$m.}
\label{fig:amplhist}
\end{center}
\end{figure}

A preliminary calculation of the electric field to simulate the effect of
surface roughness in bakelite RPC has been done as a proof of concept.
The surface profile of the bakelite sheet (Grade P120) has been studied using a DEKTAK 117 profilometer
which sampled 2500 times over a small span of 5 cm at three arbitrary locations.
Fig.~\ref{fig:linfit} shows the measurement taken
at one of these locations where a least square fit has been done in order to define the baseline of
the surface. In Fig.~\ref{fig:surfacedata} the residuals of the data with respect to the baseline are depicted as well to indicate the
amplitudes of the roughness.
The amplitudes of one such measurement is represented in histogram as shown in
Fig.~\ref{fig:amplhist}. It can
be found that the maximum amplitudes of the roughness in positive (upward) and negative (downward)
directions can be of the order of 4 $\mu$m and 2 $\mu$m, respectively.
\vspace{0.5cm}

\begin{figure}[t!]
\begin{center}
\includegraphics[scale=0.4]{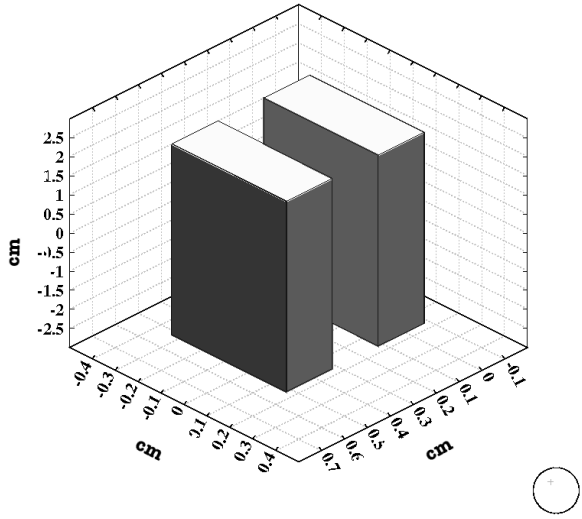}\\
\caption{\label{fig:RPCModel} The model RPC considered in the calculation.}
\end{center}
\end{figure}

For studying the effect of roughness, a device has been modeled following the geometry of an RPC.
The device comprises of two bakelite
slabs, each with thickness 2 mm (in Z-direction). A gas layer of the same height is interposed between.
A fine layer of
graphite with thickness 20 $\mu$m has been considered on other sides of the bakelites. The overall
dimension of the RPC has been taken to be 5 mm (in X-direction) $\times$ 50 mm (in Y-direction).
Fig.~\ref{fig:RPCModel} depicts the model described above.
The roughness has been added to one of the bakelite surface by building up a tiny structure with
dimensions corroborated by the measurements. Two such structures
have been introduced to represent the surface roughness with positive and negative amplitudes.
For the preliminary calculation, the shape of the roughness has been considered as a rectangular ridge
stretched along the whole length of the device in Y-direction, i.e. 50 mm, with height either
4 $\mu$m (in positive Z-direction) or 2 $\mu$m (in negative Z-direction). The cross-section
of the base of the ridge has been taken to be 50 $\mu$m $\times$ 50 mm.

\begin{figure}[htb]
\begin{center}
\includegraphics[scale=0.65]{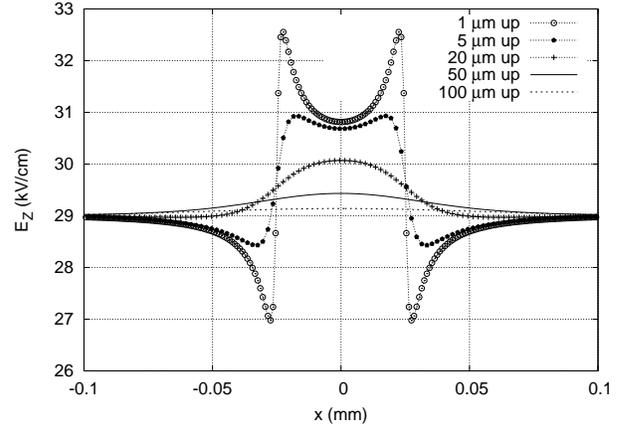}\\
\caption{\label{fig:FieldPosRough} Variation of electric field with distance away from the positive
roughness.}
\end{center}
\end{figure}

The electric field (Z-component) has been evaluated at several positions with respect to the top edge of
the structure for the
positive roughness as shown in Fig.~\ref{fig:FieldPosRough}. It can be noticed that a change of about
12\% in the field value is possible within 1 $\mu$m from the structure. The distortion in the field
due to the structure diminishes with the distance and it takes more than 100 $\mu$m to get nullified.

\begin{figure}[htb]
\begin{center}
\includegraphics[scale=0.65]{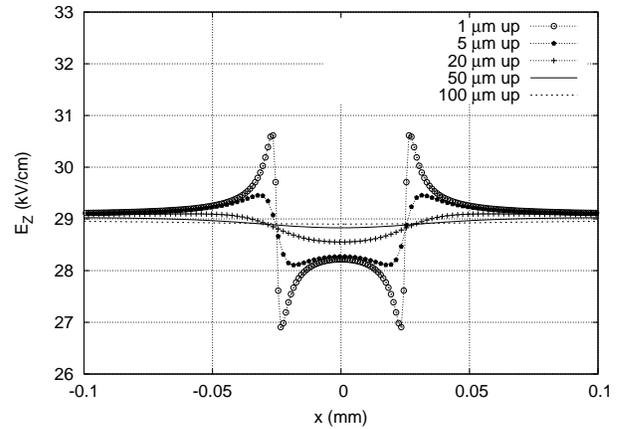}\\
\caption{\label{fig:FieldNegRough} Variation of electric field with distance near the negative
roughness.}
\end{center}
\end{figure}

The same calculation has been carried out for the model with negative roughness. The results
are illustrated in Fig.~\ref{fig:FieldNegRough}.
Here the maximum change is about 5\% at the closest vicinity of 1 $\mu$m from the bakelite surface which
is 3 $\mu$m away from the base of the roughness ridge. It may inflate upto approximately 7-8\% when
studied at just 1 $\mu $m above the base. Here, obviously the effect of the roughness dies down within
100 $\mu $m from the bakelite surface.

\section{Conclusions and outlook}
\label{}
Systematic studies on the performances of the silicone
coated bakelite RPCs were performed both in the streamer and the avalanche mode.
An efficiency >~90\% and time resolution $\sim$ 2 ns (FWHM) were obtained
in both the modes of operation. The measured induced charges on the RPC pick-up strips were found to be 1-2 pC in the avalanche mode, and $\sim$ 100 pC in case of the streamer mode of operation.
The preliminary calculation proves that the minute structures with straightforward shape on the
bakelite surface do affect the field configuration by 5 - 12 \% depending on its height or depth
in its close proximity. It indicates that the structures with different shapes and sizes may be
significant in distorting the nearby field configuration. A detailed study on the roughness
structures is therefore required, corroborated by precise measurements to advance the investigation on the effect of roughness on the performance of the RPC.

\section{Acknowledgement}
\label{}
We acknowledge the service rendered by Mr. Ganesh Das of
VECC for meticulously fabricating the detectors and Mr. Chandranath
Marick of SINP for making some of the related electronic circuits.


\noindent
\begin{thebibliography}{50}
\bibitem{RSRC81}R. Santonico, R. Cardarelli, Nucl. Inst. and Meth. 187 (1981) 377.
\bibitem{SB109}S.Biswas, et al., Nucl. Instr. and Meth. A 602 (2009) 749.
\bibitem{SB309}S.Biswas, et al., Nucl. Instr. and Meth. A 617 (2010) 138.
\bibitem{SB209}S.Biswas, et al., Nucl. Instr. and Meth. A 604 (2009) 310.
\bibitem{INO06}INO Project Report, INO/2006/01, June 2006, $\langle$http://www.imsc.res.in/$\sim$ino/$\rangle$.
\bibitem{CLu09}Changguo Lu, Nucl. Instr. and Meth. A 602 (2009) 761.
\bibitem{Garfield}http://garfield.web.cern.ch
\bibitem{neBEM}http://nebem.web.cern.ch
\bibitem{SBose109}S.Bose, et al., Nucl. Instr. and Meth. A 602 (2009) 839.
\bibitem{CLu06}C. Lu, SNIC Symposium, Stanford, California -3-6 April 2006.
\end{thebibliography}
\end{document}